\documentclass[prd,aps,preprint,amsmath,amssymb,eshowkeys,nofootinbib]{revtex4-1}
\usepackage[mathscr]{euscript}
\usepackage{dcolumn}
\usepackage{multirow}
\usepackage{bm}
\usepackage{graphicx}
\usepackage{subfigure}
\usepackage{slashbox}
\usepackage{rotating}
\usepackage{amsmath,amssymb,amsthm}
\usepackage[colorlinks=true,linkcolor=red,urlcolor=green,citecolor=green]{hyperref}
\newcommand{\bea}{\begin{eqnarray}}
\newcommand{\eea}{\end{eqnarray}}
\newcommand{\beq}{\begin{equation}}
\newcommand{\eeq}{\end{equation}}

\def\/{\over}

\begin{document}

\title{Evolution of the early universe in Einstein-Cartan theory}
\author{Qihong Huang$^{1}$\footnote{Corresponding author: huangqihongzynu@163.com}, He Huang$^{2,3}$, Bing Xu$^{4}$, and Kaituo Zhang$^{5}$}
\affiliation{
$^1$ School of Physics and Electronic Science, Zunyi Normal University, Zunyi, Guizhou 563006, China\\
$^2$ College of Mechanical and Electrical Engineering, Jiaxing Nanhu University, Jiaxing, Zhejiang 314001, China\\
$^3$ Institute of Applied Mechanics, Zhejiang University, Hangzhou, Zhejiang 310058, China\\
$^4$ School of Electrical and Electronic Engineering, Anhui Science and Technology University, Bengbu, Anhui 233030, China\\
$^5$ Department of Physics, Anhui Normal University, Wuhu, Anhui 241000, China
}

\begin{abstract}
Einstein--Cartan theory is a generalization of general relativity that introduces spacetime torsion. In this paper, we perform phase space analysis to investigate the evolution of the early universe in Einstein--Cartan theory. By studying the stability of critical points in the dynamical system, we find that there exist two stable critical points which represent an Einstein static solution and an expanding solution, respectively. After analyzing the phase diagram of the dynamical system, we find that the early universe may exhibit an Einstein static state, an oscillating state, or a bouncing state. By assuming the equation of state $\omega$ can decrease over time $t$, the universe can depart from the initial Einstein static state, oscillating state, or bouncing state and then evolve into an inflationary phase. Then, we analyze four different inflationary evolution cases in Einstein--Cartan theory and find that a time-variable equation of state $\omega$ cannot yield values of $n_{s}$ and $r$ consistent with observations, while a time-invariant equation of state $\omega$ is supported by the Planck 2018 results. Thus, in Einstein--Cartan theory, the universe likely originates from a bouncing state rather than an Einstein static state or an oscillating state.
\end{abstract}

\maketitle

\section{Introduction}

Based on general relativity, the standard cosmological model was established and can be used to describe the evolution of the universe. Although the standard cosmological model has achieved great success, it implies that the universe originated from a big bang singularity. However, in general relativity, spacetime is considered as a Riemannian manifold with vanishing torsion and zero non-metricity. The generalization of general relativity in a spacetime with torsion is named Einstein--Cartan theory, where the torsion arises from the intrinsic spin of elementary particles. Since there is a relationship between the intrinsic spin of fermionic matter and spacetime torsion, spacetime torsion is not a dynamical quantity~\cite{Hehl1976, Hehl1985, Hehl1995}. Thus, the spinor field can be incorporated into torsion-free general relativity~\cite{Lledo2010}, and Einstein--Cartan theory can be equivalent to general relativity with an effective perfect fluid in the energy-momentum tensor~\cite{Weyssenhoff1947, Obukhov1987, Smalley1994, de Berredo-Peixoto2009, Vakili2013}, which can behave as stiff matter with negative energy density in cosmology~\cite{Hehl1974, Nurgaliev1983, Gasperini1986}. Since this negative energy density directly leads to gravitational repulsion, it became very significant in the early universe. Therefore, the big bang singularity in the standard cosmological model can be resolved in Einstein--Cartan theory through a nonsingular big bounce~\cite{Poplawski2012, Unger2019}, leading to Einstein static universe~\cite{Atazadeh2014} or emergent universe models~\cite{HuangQ2015}. Furthermore, the Einstein--Cartan theory not only can address the flatness and horizon problems without requiring inflation~\cite{Poplawski2010, Poplawski2011, Poplawski2012a}, but also leads to inflation~\cite{Gasperini1986} and late-time acceleration~\cite{Shie2008}. Recently, Einstein--Cartan theory has regained much attention and has been widely studied in inflationary models~\cite{Marco2024, He2024a, He2024, Piani2022, Shaposhnikov2021}, preheating~\cite{Piani2023}, quantum cosmology~\cite{Brandt2024, Isichei2023}, gravitational waves~\cite{Ranjbar2024, Elizalde2023, Battista2022, Battista2021}, Hubble tension~\cite{Akhshabi2023}, Morris--Thorne wormhole~\cite{Soni2023}, gravitational collapse~\cite{Hensh2021}, and other physical effects~\cite{Costa2024, Falco2024, Falco2023, Battista2023, Bondarenko2021}.

Phase space analysis is a dynamical method to analyze the qualitative behavior of a dynamical system. In this approach, critical points obtained from the solutions of the autonomous system can be used to describe the evolution of the system. The stable critical points are referred to as attractors, which describe the final state of the system. When applied to cosmology, it can be used to analyze the late-time evolution of the universe and has been extensively studied in many cosmological models, such as single scalar field models~\cite{Roy2015, Dutta2016, Bhatia2017, Sola2017}, $f(R)$ gravity~\cite{Guo2013}, $f(T)$ theory~\cite{Wu2010, Wei2012}, mimetic gravity~\cite{Dutta2018}, the Chaplygin model~\cite{HuangQ2021a}, holographic dark energy~\cite{Setare2009, Banerjee2015, Huang2019, Bargach2019, HuangQ2021, HuangH2021, HuangH2022}, and so on. Recently, phase space analysis has been extended to analyze the evolution of the early universe by defining some new dimensionless variables, which can describe the expansion or contraction of the universe~\cite{Millano2023}. So, can this method be used to answer the question of whether the early universe originated from an Einstein static state, an oscillating state, or a bouncing state within the framework of Einstein--Cartan theory?

Inflation is a period of exponential expansion before the radiation-dominated era in modern cosmology~\cite{Guth1981, Linde1982}. It not only addresses the challenges of the Big Bang cosmology but also explains the quantum origin of the Cosmic Microwave Background temperature anisotropies and the Large-Scale Structure~\cite{Mukhanov1981, Lewis2000, Bernardeau2002}. During inflation, the small quantum fluctuations are amplified to physical scales and lead to nearly scale-invariant, Gaussian, and adiabatic primordial perturbations~\cite{Weinberg2008}. This information is encoded in the primordial scalar power spectrum, which is characterized by the scalar spectral index $n_{s}$ and constrained by the Planck 2018 results as $n_{s}=0.9668 \pm 0.0037$~\cite{Planck2020}. Based on the constraint on the scalar spectral index $n_{s}$, various inflation models have been proposed~\cite{Ding2024, Ragavendra2024, Pozdeeva2024, Zhang2024, Lambiase2023, Afshar2023, Bhat2023, Dioguardi2022, Karciauskas2022, Chen2022, Capozziello2021, Forconi2021, Cai2021, Gamonal2021, Fu2020, Akin2020, Fu2019, Granda2019, Gonzalez-Espinoza2019, Granda2019a, Yi2018, Casadio2018, Odintsov2018, Tahmasebzadeh2016, Yang2015, Koh2014, Gao2014, Antusch2014, Guo2010, Satoh2010, Kaneda2010, Huang2025, Gron2022}. So, in the Einstein--Cartan theory, after the universe originates from an Einstein static state, an oscillating state, or a bouncing state and then evolves into the inflation era, can it yield $n_s$ consistent with observations?

This paper has two objectives: to examine which states may have existed in the early universe, and to analyze inflation under observational constraints in Einstein--Cartan theory. The paper is organized as follows. In Section \ref{sec2}, we briefly review the field equations in Einstein--Cartan theory. In Section \ref{sec3}, we analyze the evolution of the early universe in Einstein--Cartan theory. In Section \ref{sec4}, we analyze the inflation in Einstein--Cartan theory. Finally, our main conclusions are presented in Section \ref{sec5}.

\section{Field Equation}
\label{sec2}
In Einstein--Cartan theory, the field equation can be written as~\cite{Gasperini1986, Smalley1994}
\beq\label{G0}
G^{\mu\nu}=\kappa \widetilde{T}^{\mu\nu},
\eeq
with
\beq
\widetilde{T}^{\mu\nu}=T^{\mu\nu}+\theta^{\mu\nu}=(\rho+p-\rho_{s}-p_{s})u^{\mu}u^{\nu}-(p-p_{s})g^{\mu\nu},
\eeq
and
\beq
\rho_{s}=p_{s}=\frac{\kappa}{4} \sigma^{2}.
\eeq
Here, %MDPI: Please confirm whether the paragraph format should be kept without indentation, please check the whole text, no more highlights below.
 $u^{\mu}$ is the four-velocity, $\sigma^{2}$ represents the spin density scalar, and $\kappa=8\pi G$. Since $p_{s}=\rho_{s}$, the effect of torsion and spin matter can be treated as stiff matter with negative energy density and pressure.

To study the evolution of the universe in Einstein--Cartan theory, we consider a homogeneous and isotropic universe described by the Friedman--Lema$\hat{i}$tre--Robertson--Walker (FLRW) spacetime with the metric
\beq
ds^2=dt^2-a^2(t) \bigg[\frac{dr^2}{1-k r^2}+r^2(d\theta^2+\sin^2\theta d\phi^2)\bigg],
\eeq
where $t$ is the cosmic time, $a(t)$ represents the cosmic scale factor, and $k=0,1,-1$ denote a spatially flat, closed, or open universe, respectively. Then, substituting the metric into the field Equation~(\ref{G0}), we obtain the $(00)$ and $(ii)$ components of Equation~(\ref{G0}):
\bea
&& H^{2}+\frac{k}{a^{2}}=\frac{\kappa}{3}(\rho-\rho_{s}),\label{F0}\\
&& 2\dot{H}+3H^{2}+\frac{k}{a^{2}}=-\kappa (p-p_{s}),\label{Fi}
\eea
in which $\rho=\frac{1}{2}\dot{\phi}^{2}+V$ and $p=\frac{1}{2}\dot{\phi}^{2}-V$ are the energy density and the pressure of the scalar field $\phi$, and $\rho$ and $\rho_{s}$ satisfy the continuity equations
\beq
\dot{\rho}+3H(1+\omega)\rho=0, \qquad \dot{\rho_{s}}+6H \rho_{s}=0.
\eeq
Here, $\omega$ is the equation of state and satisfies $-1 \leq \omega \leq 1$.

\section{Phase Space Analysis}
\label{sec3}
To analyze the dynamical evolution of the universe, we introduce the following dimensionless variables~\cite{Millano2023}:
\beq\label{dv}
\Omega=\frac{\kappa}{3}R^{2}\rho, \qquad \Omega_{s}=\frac{\kappa}{3}R^{2}\rho_{s}, \qquad \Omega_{k}=\frac{k R^{2}}{a^{2}}, \qquad Q=RH,
\eeq
where $R$ is the apparent horizon radius for the FLRW universe and it has the form
\beq
R=ar=\frac{1}{\sqrt{H^{2}+\frac{k}{a^{2}}}}.
\eeq
Using the dimensionless variables given in~(\ref{dv}), the apparent horizon radius yields the relation
\beq
Q^{2}+\Omega_{k}=1,\label{OO1}
\eeq
and the Friedmann Equation~(\ref{F0}) can be written as
\beq
Q^{2}+\Omega_{k}=\Omega-\Omega_{s},\label{OO2}
\eeq
with $-1 \leq Q \leq 1$, $0 \leq \Omega_{k} \leq 1$, $\Omega_{s} \geq 0$ and $\Omega \geq 1$. By combining Equations~(\ref{OO1}) and (\ref{OO2}), we obtain $\Omega_{s} = \Omega - 1$, showing that only one independent variable exists between $\Omega$ and $\Omega_{s}$. Then, introducing the time derivative
\beq
f'=\frac{df}{d\tau}=R \dot{f},
\eeq
we obtain the dynamical system
\bea
&& \Omega'=3(1-\omega)(1-\Omega)Q \Omega,\label{ds1}\\
&& Q'=\frac{1}{2}(1-Q^{2})[3(1-\omega)\Omega-4],\label{ds2}
\eea
defined on the phase plane with $\Omega \geq 1$ and $-1 \leq Q \leq 1$. Now, we will analyze the phase space behavior of the dynamical system~(\ref{ds1}) and~(\ref{ds2}). The critical points of the autonomous system can be obtained by setting
\beq\label{AS}
\Omega'=Q'=0.
\eeq
Then, we obtain five critical points. Since the existence conditions of these critical points are limited by $\Omega \geq 1$, $-1 \leq Q \leq 1$, and $\Omega_{s} \geq 0$, two critical points are abandoned because $\Omega_{s}=-1$. The remaining three critical points $(P_{1},P_{2},P_{3})$ are shown in Table~\ref{Tab1}. From this table, we can see that $P_{1}$ is a contracting solution, $P_{2}$ denotes an Einstein static solution, and $P_{3}$ represents an expanding solution. The existence conditions show that $P_{1}$ and $P_{3}$ always exist, whereas $P_{2}$ is determined by the equation of state $\omega$.

\begin{table}
\caption{\label{Tab1} Critical points and their stability conditions.}
\begin{center}
 \begin{tabular}{|c|c|c|c|c|c|c|}
  \hline
  \hline
  $Label$ & $(\Omega, Q)$ & $\Omega_{s}$ & $Existence$ & $Eigenvalues$ & $Conditions$ & $Points$\\
  \hline
  \multirow{2}*{${P_{1}}$}
  &\multirow{2}*{$(1,-1)$}
  &\multirow{2}*{$0$}
  &\multirow{2}*{$Always$}
  &\multirow{2}*{$-1-3\omega, 3-3\omega$}
  & $-\frac{1}{3} < \omega <1$ & $Saddle \ point$\\
  \cline{6-7}
  & & & & & $-1 \leq \omega < -\frac{1}{3}$ & $Unstable \ point$\\
  \hline
  ${P_{2}}$ & $(\frac{4}{3(1-\omega)},0)$ & $\frac{1+3\omega}{3(1-\omega)}$ & $-\frac{1}{3}\leq\omega<1$ & $-\sqrt{2(-1-3\omega)}, \sqrt{2(-1-3\omega)}$ & $-\frac{1}{3} < \omega < 1$ & $Center \ point$\\
  \hline
  \multirow{2}*{${P_{3}}$}
  &\multirow{2}*{$(1,1)$}
  &\multirow{2}*{$0$}
  &\multirow{2}*{$Always$}
  &\multirow{2}*{$1+3\omega,-3+3\omega$}
  & $-\frac{1}{3} < \omega < 1$ & $Saddle \ point$\\
  \cline{6-7}
  & & & & & $-1 \leq \omega < -\frac{1}{3}$ & $Stable \ point$\\
  \hline
  \hline
  \end{tabular}
\end{center}
\end{table}

To discuss the stability of the critical points, we will use the linear stability theory to analyze these points. By linearizing the autonomous system~(\ref{ds1}) and~(\ref{ds2}), we obtain two differential equations. The stabilities of these critical points are fully determined by the eigenvalues of the coefficient matrix of the two differential equations. If all eigenvalues of the critical point possess negative real parts, the point is stable; if all eigenvalues have a positive real part, the point is unstable; if at least two eigenvalues have real parts with opposite signs, this point is called a saddle point. In addition, if the eigenvalue has a zero real part, the critical point is called a non-hyperbolic point, for which the stability of the critical point cannot be determined by linear stability theory. To analyze the stability of the non-hyperbolic point, center manifold theory~\cite{Boehmer2012, Bargach2019} or the numerical method~\cite{Bargach2019, Dutta2016, Dutta2017, Dutta2019} can be used.

\begin{figure*}[htp]
\begin{center}
\includegraphics[width=0.45\textwidth]{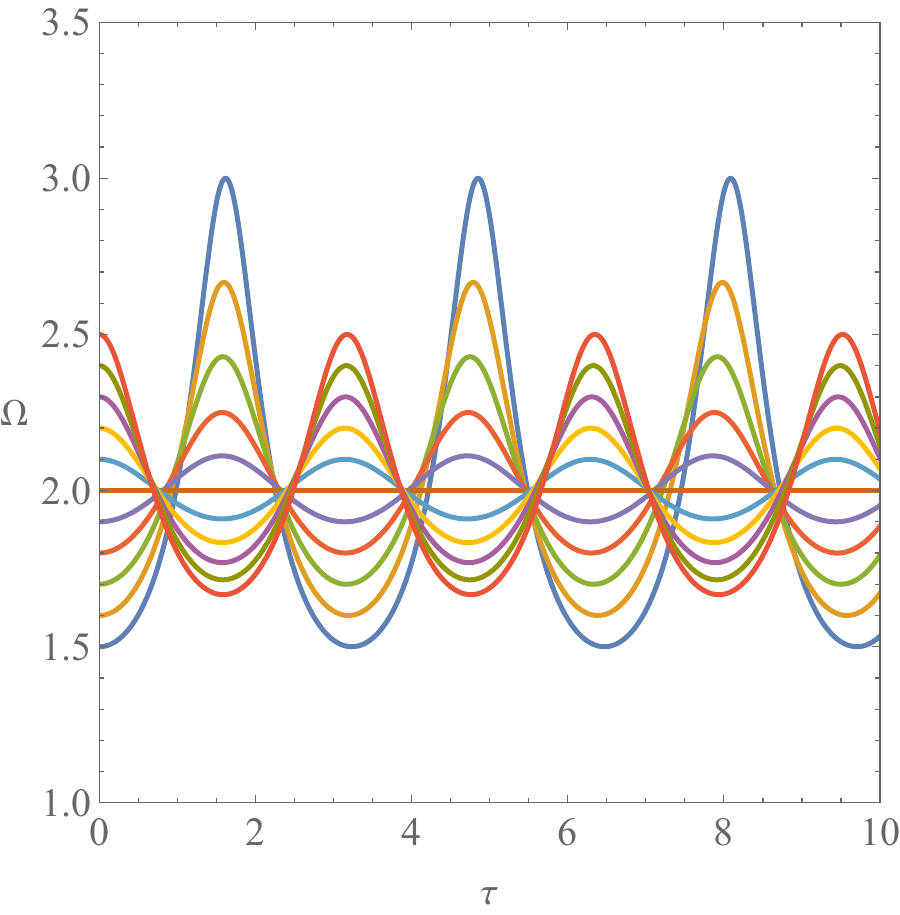}
\includegraphics[width=0.46\textwidth]{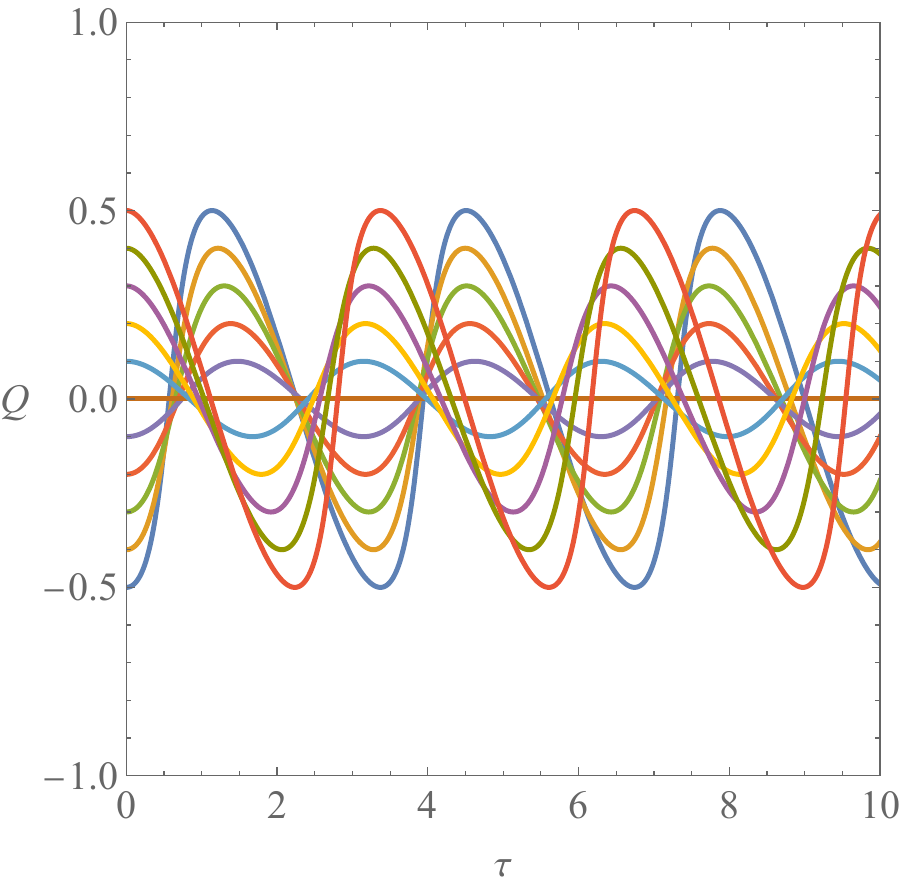}
\caption{\label{Fig1} Projections of the time evolution of phase space trajectories along the $\Omega$-axis and $Q$-axis, which determine the stability of point $P_{2}$. These panels are plotted for $\omega=\frac{1}{3}$.}
\end{center}
\end{figure*}

After some calculations, we obtain the eigenvalues and stability conditions of these critical points, which are shown in Table~\ref{Tab1}. Specially, for the critical point $P_{2}$, the eigenvalues are purely imaginary, and this critical point is called a center, which is a stable point~\cite{Brannan2015}. Projections of the time evolution of phase space trajectories for $P_{2}$ are shown in Figure~\ref{Fig1}, which indicates that a perturbation from the critical point will lead to an oscillation around this point rather than an exponential deviation. According to the results shown in Table~\ref{Tab1}, we can analyze the evolution of the early universe according to the equation of state $\omega$.

\begin{figure*}[htp]
\begin{center}
\includegraphics[width=0.45\textwidth]{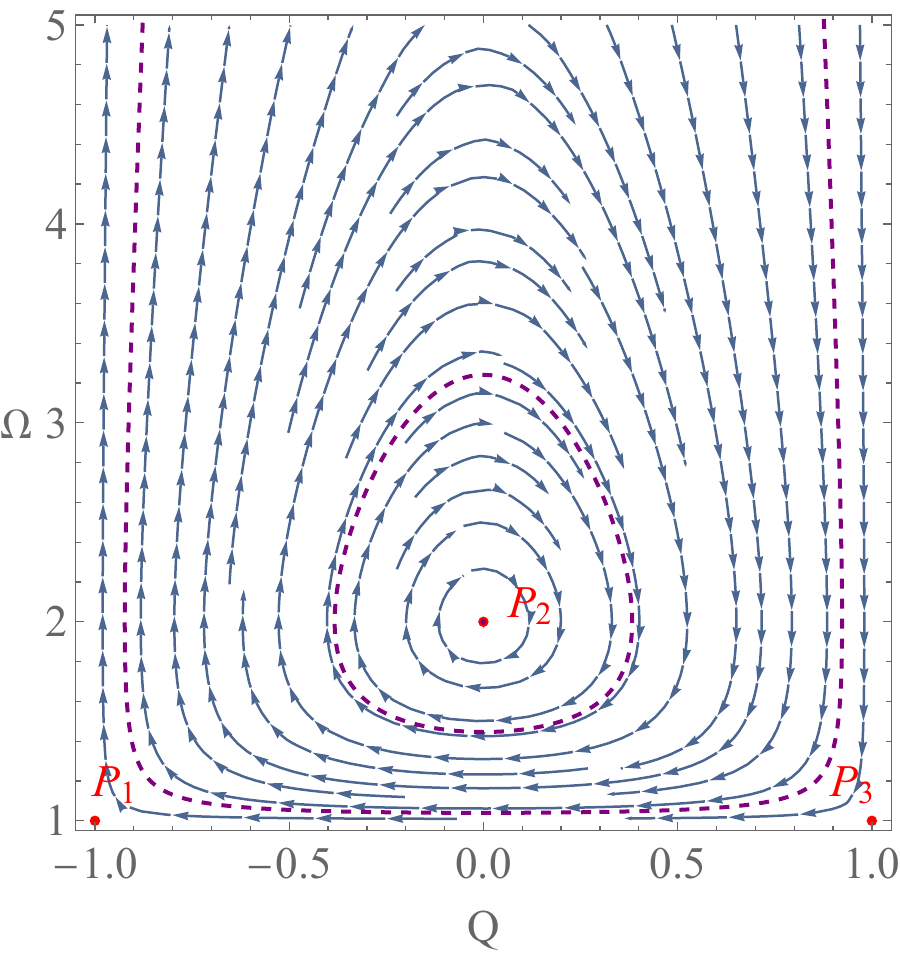}
\includegraphics[width=0.45\textwidth]{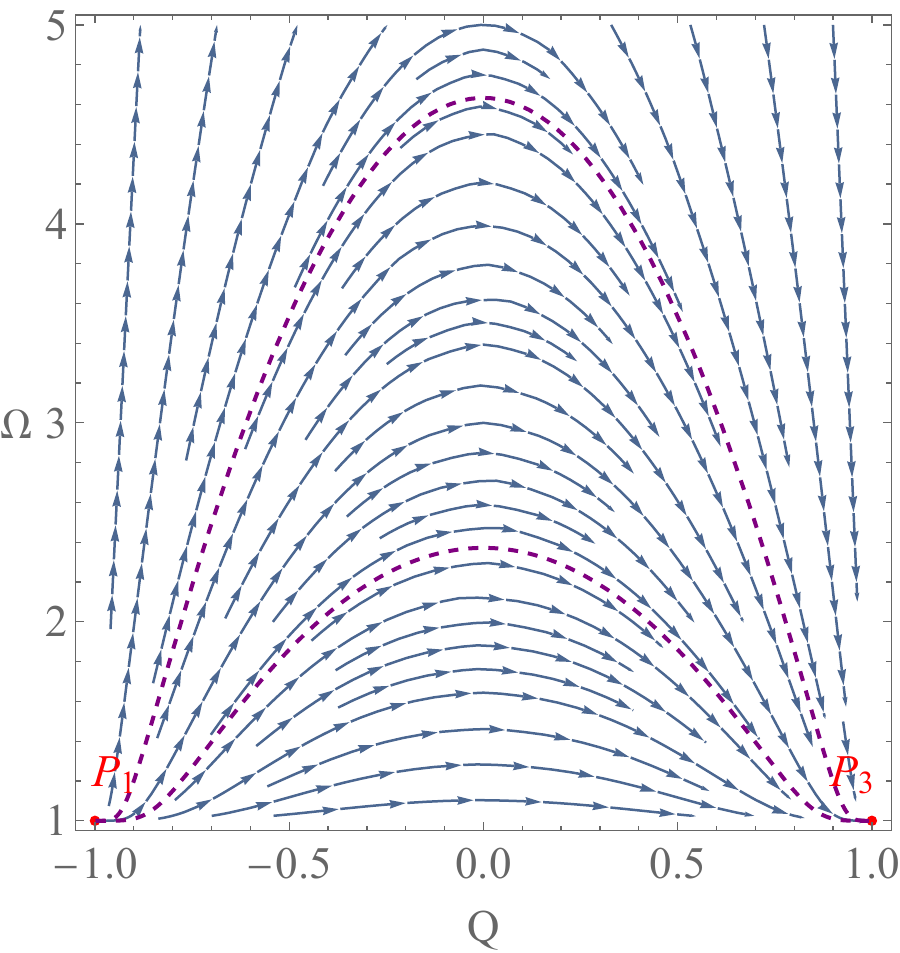}
\caption{\label{Fig2} Phase diagram of $(\Omega,Q)$. These panels are plotted for $\omega=\frac{1}{3}$ and $-\frac{2}{3}$, respectively. The red points denote the critical points, while the purple dashed lines represent examples of the evolutionary curves of the early universe.}
\end{center}
\end{figure*}

For $-\frac{1}{3} < \omega <1$, there are three critical points $P_{1}$, $P_{2}$, and $P_{3}$ in the autonomous system. $P_{1}$ denotes a saddle point corresponding to a contracting solution, $P_{2}$ denotes a stable Einstein static solution, and $P_{3}$ represents a saddle point corresponding to an expanding solution. In this case, the universe can become either an Einstein static universe or an oscillating universe. When $(\Omega, Q)$ takes the value of $P_{2}$, the early universe is an Einstein static state, while it becomes an oscillating state if $(\Omega, Q)$ takes other values. These cases are depicted in the left panel of Figure~\ref{Fig2}. In these figures, the red points denote the critical points, and the purple dashed lines represent an example of the evolutionary curves of the early universe.

For $-1 \leq \omega < -\frac{1}{3}$, the autonomous system has two critical points $P_{1}$ and $P_{3}$. $P_{1}$ denotes an unstable contracting solution, while $P_{3}$ denotes a stable expanding solution. Therefore, in this case, the universe will evolve from a contraction phase to an expanding phase, and it represents a bouncing universe, as shown in the right panel of Figure~\ref{Fig2}.

Thus, the evolution of the early universe in Einstein--Cartan theory is determined by the initial conditions and the equation of state $\omega$, and the initial state of the %MDPI: Please confirm if the bold is unnecessary and can be removed. The following highlights are the same.
universe may be an Einstein static universe, an oscillating universe, or a bouncing universe. After the universe originates from one of these initial states, it can evolve into a subsequent inflation era since point $P_{3}$ is an attractor for $-1 \leq \omega < -\frac{1}{3}$.

In contrast to %MDPI: We have removed the bold formatting of this paragraph. Please confirm..
analyses of the Einstein static state in Refs.~\cite{Atazadeh2014,HuangQ2015} where the static state requires $\dot{a}=\ddot{a}=0$, and investigations of the bouncing state in Refs.~\cite{Poplawski2012, Unger2019}, which impose $\dot{a}=0$ at the bounce, our analysis adopts phase space analysis without invoking these conditions.

\begin{figure*}[htp]
\begin{center}
\includegraphics[width=0.45\textwidth]{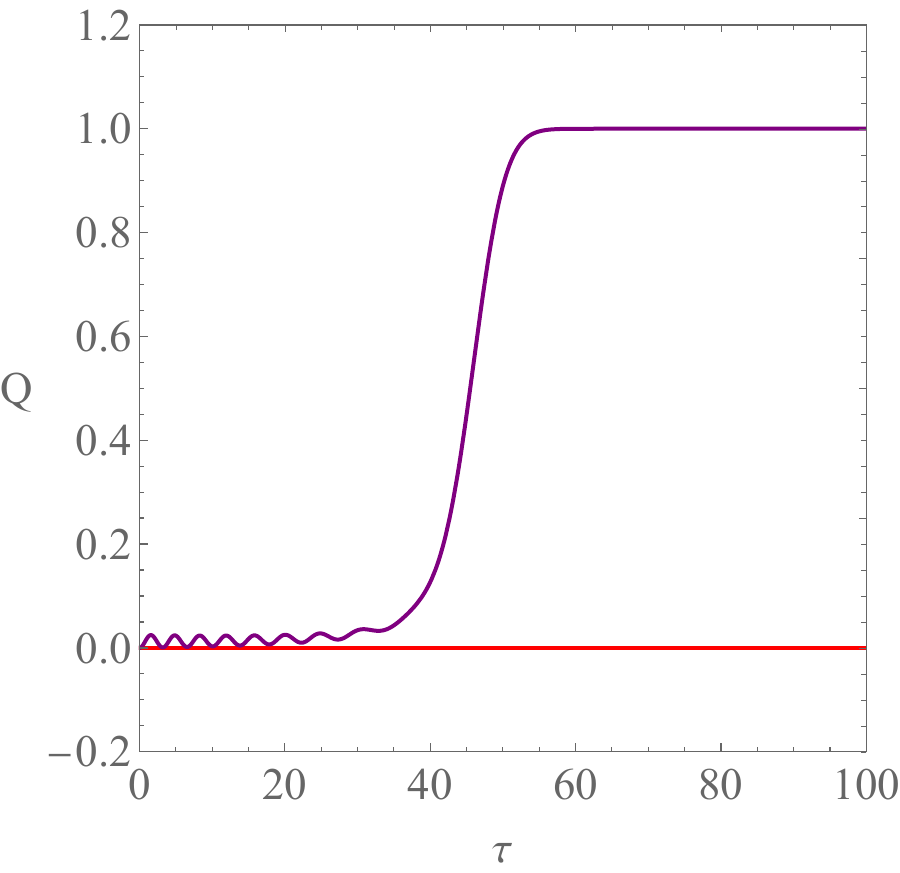}
\includegraphics[width=0.45\textwidth]{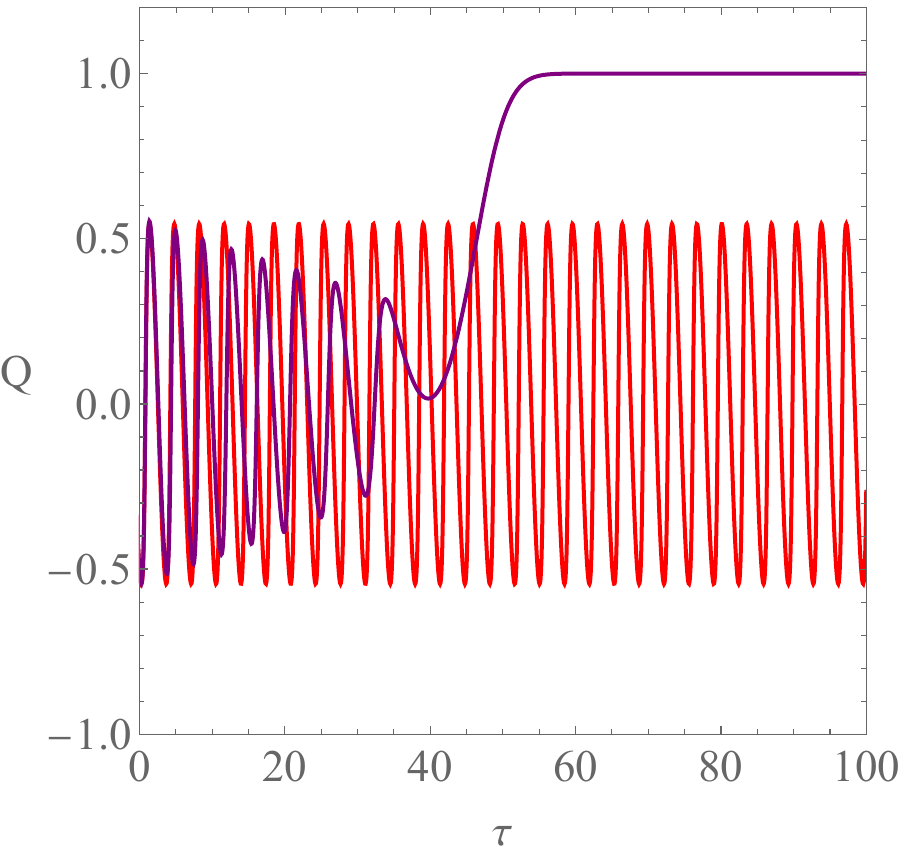}
\caption{\label{Fig3} Evolutionary curve of $Q$ with time $\tau$. The red lines are plotted for $\omega=\frac{1}{3}$ with different initial values, while the purple lines are plotted for a time-variable $\omega$. The red line in the left panel denotes an Einstein static universe, while the one in the right panel represents an oscillating universe.}
\end{center}
\end{figure*}

It is interesting to note that the stability conditions of $P_{2}$ have also been obtained in the Einstein static universe~\cite{Atazadeh2014} and the emergent universe~\cite{HuangQ2015}. When the early universe is an Einstein static universe and $\omega$ decreases to less than $-\frac{1}{3}$, the stability condition of the Einstein static solution $P_{2}$ is broken, and $P_{3}$ becomes a stable point, acting as an attractor. As a result, the universe exits from the stable Einstein static state $P_{2}$ and evolves into an expanding state described by $P_{3}$. In the left panel of Figure~\ref{Fig3}, we have plotted this transition. In this figure, the red line denotes an Einstein static universe with $\omega=\frac{1}{3}$, $\Omega=2$, and $Q=0$, while the purple line illustrates this transition by considering $\omega$ decreasing over time $\tau$. We can see that when $\tau$ increases to a critical value, the universe transitions from the Einstein static phase to an expanding phase with $Q=1$, which is depicted by $P_{3}$.

If the initial state of the universe is an oscillating universe and $\omega$ decreases to less than $-\frac{1}{3}$, breaking down the oscillating conditions, the universe can exit from the oscillating state and evolve into an expanding phase described by $P_{3}$. In the right panel of Figure~\ref{Fig3}, we have plotted this evolutionary process. In this figure, the red line denotes an oscillating universe with $\omega=\frac{1}{3}$, while the purple line represents the evolutionary curve with a time-variable $\omega$ that decreases with time $\tau$. We can see that when $\tau$ increases to a critical value, the universe departs from the oscillating phase and evolves into an expanding phase with $Q=1$.

\section{Inflation}
\label{sec4}
In the previous section, we find that in Einstein--Cartan theory, the early universe can originate from three different states: (i) an Einstein static state; (ii) an oscillating state; (iii) a bouncing state. After the universe originates from one of these initial states, it can evolve into a subsequent inflation era. In this section, we analyze the scalar spectral index $n_{s}$ and the tensor-to-scalar ratio $r$ produced during the universe's evolution from each of the three different initial states into the inflationary epoch, and then use Planck 2018 results to constrain them and discuss from which initial state the universe is likely to originate.

To analyze inflation in Einstein--Cartan theory, we ignore the role of spatial curvature $k$ and the spin density scalar $\sigma^{2}$ in the Mukhanov--Sasaki equation and focus on their effect on the dynamical evolution process, as spacetime is nearly flat and the scale factor $a$ is very large after inflation ends. Thus, the inflation is driven by the scalar field. Then, we adopt the slow-roll parameter $\epsilon_{n}$, defined by the Hubble parameters $H$ and its derivative $\dot{H}$, which are given as~\cite{Martin2014}
\beq
\epsilon_{1}=-\frac{\dot{H}}{H^{2}}, \qquad \epsilon_{2}=\frac{\dot{\epsilon_{1}}}{H \epsilon_{1}}.
\eeq
Combining Equations~(\ref{F0}) and~(\ref{Fi}), we obtain
\beq\label{dH}
\dot{H}=-3H^{2}+\frac{\kappa}{2}(1-\omega)\rho-2\frac{k}{a^{2}}.
\eeq
Using Equations~(\ref{F0}) and~(\ref{dH}), we can write the slow-roll parameters $\epsilon_{1}$ and $\epsilon_{2}$ as
\bea
&& \epsilon_{1}=\frac{6\frac{k}{a^{2}}+3\kappa[2\rho_{s}-(1+\omega)\rho]}{6\frac{k}{a^{2}}+2\kappa(\rho_{s}-\rho)},\label{e1}\\
&& \epsilon_{2}=\frac{\kappa\Big[(1+3\omega)^{2} \frac{k}{a^{2}} \rho+[-16\frac{k}{a^{2}}+3\kappa(1-\omega)^{2}\rho]\rho_{s}\Big]}{\Big[3\frac{k}{a^{2}}+\kappa(\rho_{s}-\rho)\Big]\Big[2\frac{k}{a^{2}}+\kappa[2\rho_{s}-(1+\omega)\rho]\Big]}.\label{e2}
\eea
Now, utilizing the slow-roll parameters $\epsilon_{1}$ and $\epsilon_{2}$, we can calculate the scalar spectral index $n_{s}$ and the tensor-to-scalar ratio $r$ using the following expressions~\cite{Martin2014}:
\bea
&& n_{s}=1-2\epsilon_{1}-2\epsilon_{2},\label{ns0}\\
&& r=16\epsilon_{1}.\label{r0}
\eea

\subsection{$k=0$}

\begin{figure*}[htp]
\begin{center}
\includegraphics[width=0.445\textwidth]{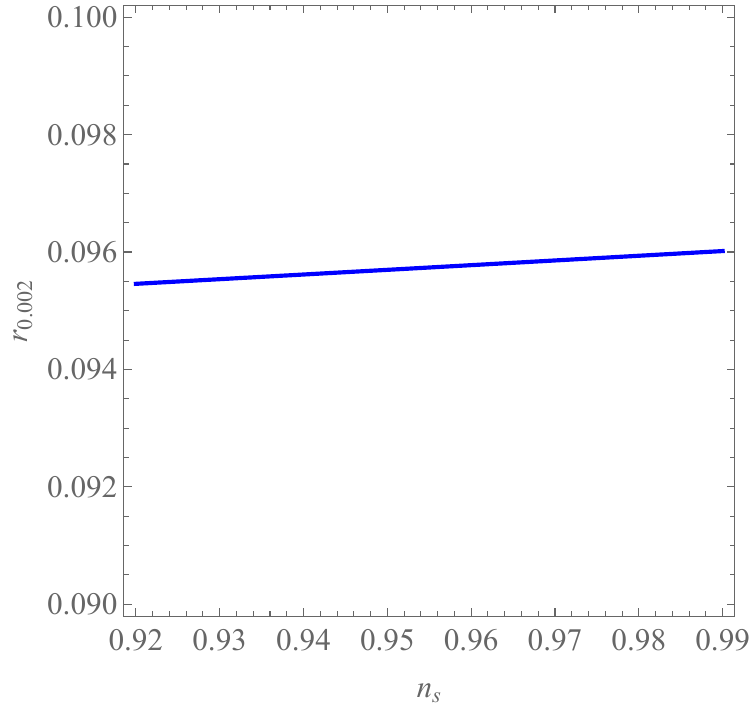}
\includegraphics[width=0.45\textwidth]{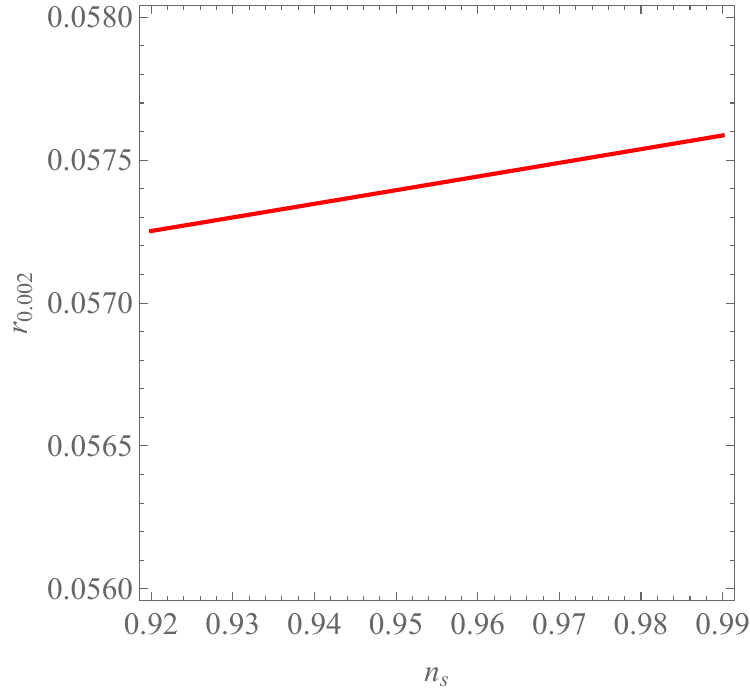}
\includegraphics[width=0.45\textwidth]{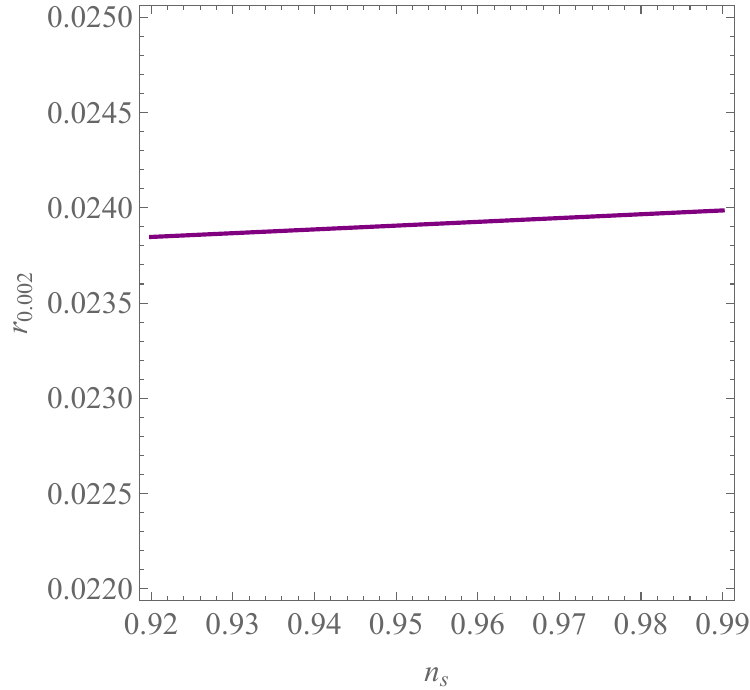}
\includegraphics[width=0.45\textwidth]{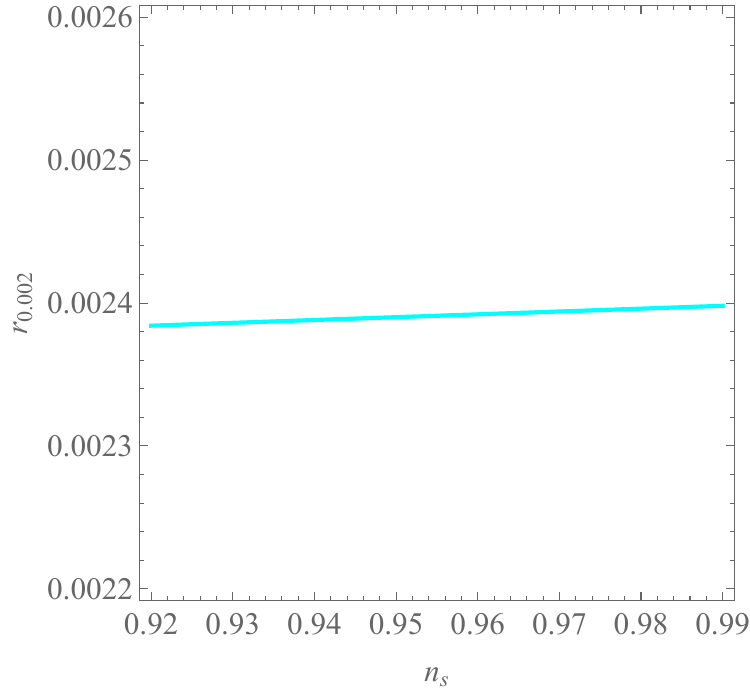}
\caption{\label{Fig4} Relation between $n_{s}$ and $r_{0.002}$ for $\omega=-0.996$,$-0.9976$,$-0.9990$, and $-0.9999$, respectively.}
\end{center}
\end{figure*}

Considering the situation where the spacetime is nearly flat after inflation ends, we assume $k=0$, and Equation~(\ref{e1}) provides the relation
\beq
\frac{\rho_{s}}{\rho} = 1-\frac{3(1-\omega)}{2(3-\epsilon_{1})}.\label{rsr}
\eeq
Then, using Equations~(\ref{e1})--(\ref{rsr}), the scalar spectral index $n_{s}$ can be rewritten as
\beq
n_{s}=19-\frac{3}{8}r+6\omega-\frac{288(1+\omega)}{r},
\eeq
In Figure~\ref{Fig4}, considering the constraint $n_{s}=0.9668 \pm 0.0037$ for $r_{0.002}<0.058$~\cite{Planck2020}, we have plotted the relation between $n_{s}$ and $r_{0.002}$ for different $\omega$. This figure shows that to obtain $n_{s}$ and $r$ supported by Planck 2018 results, the equation of state $\omega$ needs to approach $-1$ during inflation. It is worth noting that during slow-roll inflation in General Relativity, the equation of state parameter $\omega$ satisfies $\omega=-1+\frac{2}{3}\epsilon_{1}$, which is approximately $-1$ as a result of $\dot{\phi}^{2} \ll V$~\cite{Baumann2012}. Thus, the results shown in Figure~\ref{Fig4} are consistent with the value of $\omega$ given by the slow-roll inflation in General Relativity.

\subsection{$k = 1$}

\begin{figure*}[htp]
\begin{center}
\includegraphics[width=0.55\textwidth]{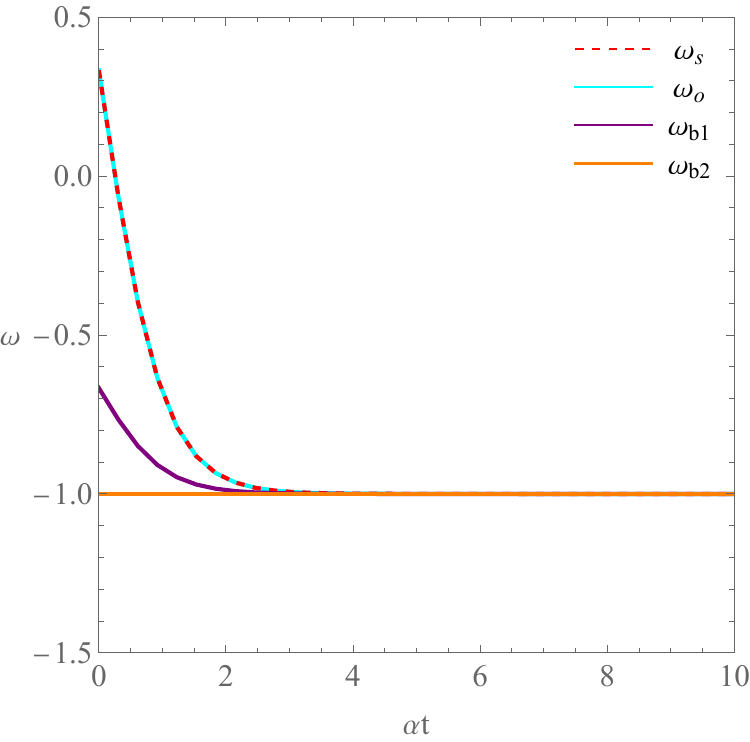}
\caption{\label{Fig5} Evolutionary curve for different $\omega$. Here, we adopt $\omega_{0} = 0.9999$.}
\end{center}
\end{figure*}

In the previous section, we found that the early universe in Einstein--Cartan theory has three different origins: (i)~Einstein static state; (ii)~oscillating state; (iii)~bouncing state. In this subsection, we will combine these results to analyze the inflation in the case $k = 1$. For the case $k = 1$, we cannot obtain the analytical expression for the relationship between $n_{s}$ and $r$. To analyze the inflation for $k = 1$, we adopt a numerical method. We analyze four different inflationary evolution cases: (i) originating from an Einstein static state with a time-variable equation of state $\omega_{s}=\frac{1}{3}[1-4\omega_{0}\tanh(\alpha t)]$, and then evolving into the inflationary phase; (ii) originating from an oscillating state with a time-variable equation of state $\omega_{o}=\frac{1}{3}[1-4\omega_{0}\tanh(\alpha t)]$, and then evolving into the inflationary phase; (iii) originating from a bouncing state with a time-variable equation of state $\omega_{b1}=-\frac{2}{3}[1+\frac{1}{2}\omega_{0}\tanh(\alpha t)]$, and then evolving into the inflationary phase; (iv) originating from a bouncing state with a time-invariant equation of state $\omega_{b2}=-\omega_{0}$, and then evolving into the inflationary phase. Parameter $\alpha$ has dimensions of inverse time, and $\omega_{0}$ is a value chosen based on the results in Figure~\ref{Fig4}. In Figure~\ref{Fig5}, we have plotted the evolutionary curves for these $\omega$. With the increase in $\alpha t$, all curves for $\omega$ approach $-\omega_{0}$, which corresponds to the equation of state during slow-roll inflation.

\begin{figure*}[htp]
\begin{center}
\includegraphics[width=0.447\textwidth]{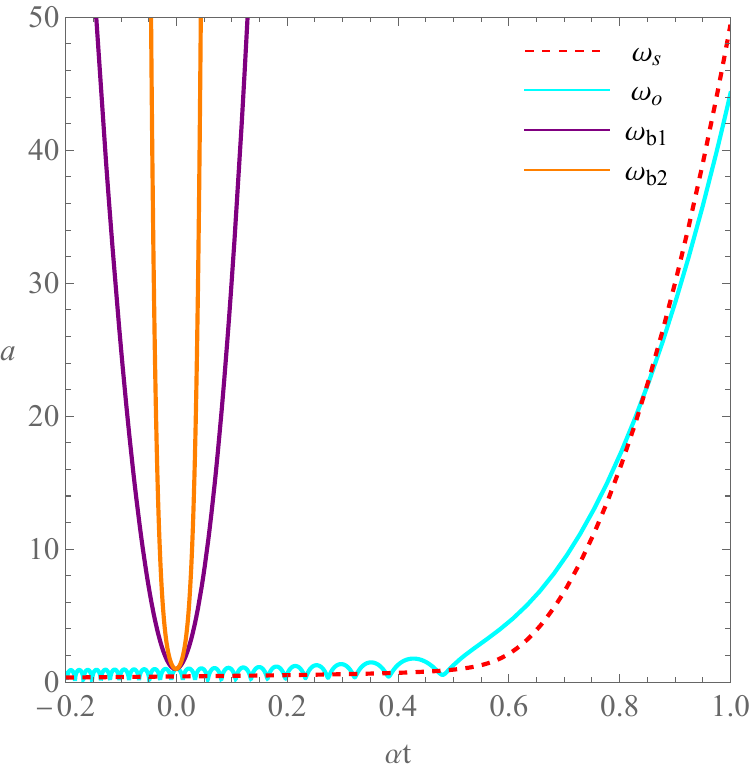}
\includegraphics[width=0.453\textwidth]{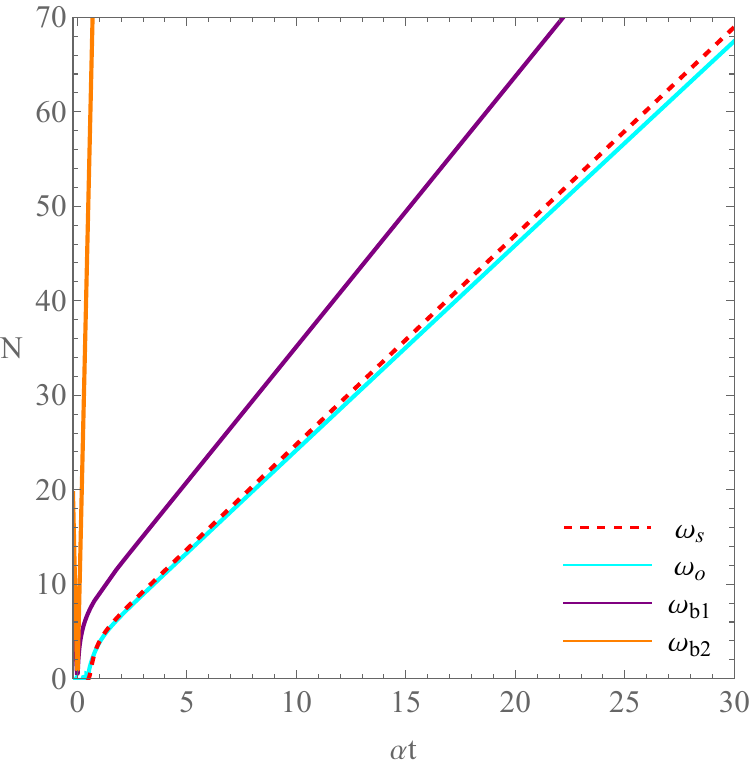}
\caption{\label{Fig6} Evolutionary curve of the scale factor $a$ and the e-folds number $N$ in different cases.}
\end{center}
\end{figure*}

Combining Equations~(\ref{F0}) and~(\ref{Fi}), we obtain the following equation by eliminating $\rho$:
\beq\label{erho}
2\dot{H}+3(\omega+1)H^{2}+(3\omega+1)\frac{1}{a^{2}}=\kappa (1-\omega)\rho_{s}.
\eeq
Solving this equation numerically for different $\omega$, respectively, we obtain different inflationary evolution curves, which are shown in Figure~\ref{Fig6}. The left panel of Figure~\ref{Fig6} shows the evolutionary curves of the scale factor $a$ over time $\alpha t$, while the right panel shows the evolutionary curves for the e-folds number $N$, which is defined as $N=\ln\frac{a}{a_{0}}$. The red dashed line indicates that the universe originates from the Einstein static state and subsequently evolves into the inflationary phase as the equation of state $\omega_{s}$ decreases; the cyan line shows that the universe starts from an oscillating state and then transitions into the inflationary phase as the equation of state $\omega_{o}$ decreases; the purple line represents that the universe begins from a bouncing state and then evolves into the inflationary phase as the equation of state $\omega_{b1}$ decreases; the orange line denotes that the universe begins from a bouncing state and then evolves into the inflationary phase with a constant equation of state $\omega_{b2}$. From these figures, we can see that inflation can occur in all these cases, and the e-folds number $N$ can exceed $65$ as time increases. Additionally, the evolutionary curves for the universe originating from a static or oscillating state are nearly overlapping.

\begin{figure*}[htp]
\begin{center}
\includegraphics[width=0.55\textwidth]{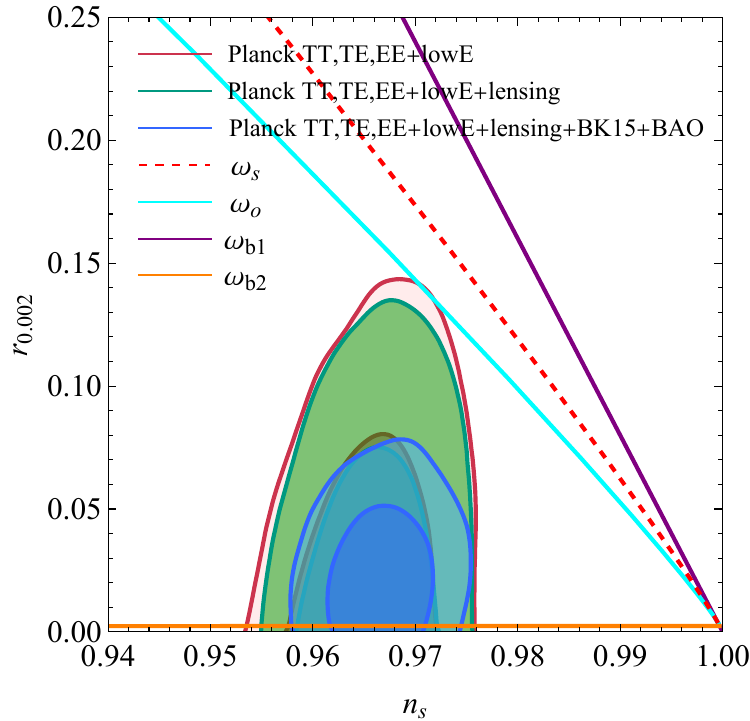}
\caption{\label{Fig7} Relation between $n_{s}$ and $r_{0.002}$ in different cases.}
\end{center}
\end{figure*}

Then, by numerically solving Equation~(\ref{erho}), we obtain the relation between the scalar spectral index $n_{s}$~(\ref{ns0}) and the tensor-to-scalar ratio $r$~(\ref{r0}) for different cases, as shown in Figure~\ref{Fig7}. In this figure, we overlay our numerical results with Planck 2018 results~\cite{Planck2020}. This figure shows that the time-variable equation of state $\omega$ cannot yield results consistent with the observations, while a time-invariant equation of state $\omega$ is supported by the Planck 2018 results, and that the scalar spectral index $n_{s}$ and tensor-to-scalar ratio $r$ depend on both the initial state of the universe and the evolution of $\omega$ during the transition to inflation. Thus, in Einstein--Cartan theory, the universe is unlikely to originate from an Einstein static state or an oscillating state since subsequent inflation cannot yield $n_{s}$ and $r$ consistent with the observations. Instead, the universe is likely to originate from a bouncing state with a time-invariant equation of state $\omega \approx -1$.

In Einstein--Cartan theory, the emergent universe cannot produce $n_{s}$ and $r$ that are consistent with observations. An emergent universe, as supported by Planck 2018 results, may exist within various modified theories including scalar-tensor theory~\cite{Miao2016,Huang2023}, mimetic gravity~\cite{HuangQ2020}, non-minimal derivative coupling models~\cite{Huang2018a,Huang2018b}, braneworld models~\cite{Zhang2016}, and f(R,T) gravity~\cite{Sharif2019}. However, additional research is required to fully explore these possibilities.

\section{Conclusions}
\label{sec5}
Einstein--Cartan theory is a generalization of general relativity that introduces spacetime torsion, which can be equivalent to general relativity with an exotic stiff perfect fluid. In this paper, we analyze the evolution of the early universe in the Einstein--Cartan theory using the phase space analysis method. We find that there are three different critical points. The stability of these critical points is determined by the equation of state $\omega$. There exist two stable critical points $P_{2}$ and $P_{3}$. $P_{3}$ denotes an expanding solution and is stable for $-1 \leq \omega < -\frac{1}{3}$, while the stable Einstein static solution $P_{2}$ requires $-\frac{1}{3}< \omega <1$ and it is a center point. Therefore, $P_{3}$ can represent the final state of the evolution of the early universe. After analyzing the phase diagram of the dynamical system and considering different initial conditions and the equation of state $\omega$, we find there may exist an Einstein static universe, an oscillating universe, or a bouncing universe in the early universe. By assuming the equation of state $\omega$ can decrease over time $t$, the universe can depart from the initial Einstein static state, oscillating state, or bouncing state and then evolve into an inflationary state.

Subsequently, we analyze the inflation in Einstein--Cartan theory. For the spatial curvature $k=0$, we find that to obtain $n_{s}$ and $r$ values supported by Planck 2018 results, the equation of state $\omega$ needs to approach $-1$ during inflation, and the value of $\omega$ is consistent with that given by the slow-roll inflation in General Relativity. For the spatial curvature $k=1$, we analyze four different inflationary evolution cases: (i) originating from an Einstein static state with a time-variable equation of state, and then evolving into the inflationary phase; (ii) originating from an oscillating state with a time-variable equation of state, and then evolving into the inflationary phase; (iii) originating from a bouncing state with a time-variable equation of state, and then evolving into the inflationary phase; (iv) originating from a bouncing state with a time-invariant equation of state, and then evolving into the inflationary phase. We find that the time-variable equation of state $\omega$ cannot yield results consistent with the observations, while a time-invariant equation of state $\omega$ is supported by the Planck 2018 results. Therefore, in Einstein--Cartan theory, the universe is unlikely to originate from an Einstein static state or an oscillating state; instead, it is likely to originate from a bouncing state with a time-invariant equation of state $\omega \approx -1$.

\begin{acknowledgments}

This work was supported by the National Natural Science Foundation of China under Grants Nos.12265019, 12405081, 11865018, 12305056, the regional first-class discipline of Guizhou province of China under Grants No. QJKYF[2018]216, the University Scientific Research Project of Anhui Province of China under Grants No. 2022AH051634.

\end{acknowledgments}

\end{document}